\documentclass[runningheads]{llncs}
\usepackage[T1]{fontenc}
\usepackage{graphicx}
\usepackage{amsmath,amssymb,amsfonts}
\usepackage{algorithmic}
\usepackage{multirow}
\usepackage{booktabs}
\usepackage{tgpagella}
\usepackage{colortbl}
\usepackage[colorlinks,urlcolor=blue,linkcolor=blue,citecolor=blue]{hyperref}
\usepackage{xcolor}
\usepackage{todonotes}

\begin{document}
\title{Expert-Guided Explainable Few-Shot Learning for Medical Image Diagnosis
}
%



\author{Ifrat Ikhtear Uddin \and
Longwei Wang \and
KC Santosh
}
%

\authorrunning{Uddin et al.}
%
\institute{AI Research Lab, Department of Computer Science, University of South Dakota, Vermillion, SD 57069, USA \\
\email{ifratikhtear.uddin@coyotes.usd.edu}\\
\email{longwei.wang@usd.edu} \\
\email{kc.santosh@usd.edu}
}
\maketitle              
%


\begin{abstract}
Medical image analysis often faces significant challenges due to limited expert-annotated data, hindering both model generalization and clinical adoption. We propose an expert-guided explainable few-shot learning framework that integrates radiologist-provided regions of interest (ROIs) into model training to simultaneously enhance classification performance and interpretability. Leveraging Grad-CAM for spatial attention supervision, we introduce an explanation loss based on Dice similarity to align model attention with diagnostically relevant regions during training. This explanation loss is jointly optimized with a standard prototypical network objective, encouraging the model to focus on clinically meaningful features even under limited data conditions. We evaluate our framework on two distinct datasets: BraTS (MRI) and VinDr-CXR (Chest X-ray), achieving significant accuracy improvements from 77.09\% to 83.61\% on BraTS and from 54.33\% to 73.29\% on VinDr-CXR compared to non-guided models. Grad-CAM visualizations further confirm that expert-guided training consistently aligns attention with diagnostic regions, improving both predictive reliability and clinical trustworthiness. Our findings demonstrate the effectiveness of incorporating expert-guided attention supervision to bridge the gap between performance and interpretability in few-shot medical image diagnosis.
\end{abstract}

\section{Introduction}

Deep learning has achieved remarkable success in medical image analysis, enabling automated disease detection, segmentation, and classification across a wide range of modalities \cite{krizhevsky2012imagenet,he2016deep,redmon2016you,wang2019representation,wang2014congestion,wang2018partial,wang2016optimization,shi2019deep,10499639,wang2024enhanced,wang2021improving}. However, these successes are often predicated on the availability of large, meticulously annotated datasets a requirement that is particularly challenging in clinical settings, where expert annotation is both time consuming and expensive. In many practical scenarios, only a few labeled examples per diagnostic category may be available, limiting the applicability of conventional supervised deep learning approaches. 

Few-shot learning (FSL) aims to address this limitation by enabling models to generalize from a limited number of labeled samples \cite{liu2023overcoming,hersche2022constrained}. Despite recent progress, few-shot models often remain susceptible to overfitting and lack transparency in their decision-making processes, raising concerns for deployment in safety-critical clinical applications where interpretability is essential.

Simultaneously, explainable artificial intelligence (XAI) \cite{selvaraju2017grad,lundberg2017unified,ribeiro2016should} has emerged as a key research direction to enhance model transparency and facilitate clinician trust \cite{mahamud2025enhancing,ranabhat2025multi}. Methods such as Grad-CAM provide post-hoc visual explanations by highlighting image regions that contribute to model predictions \cite{wang2021explaining}. However, these explanations are typically generated after training and do not influence the learning process itself \cite{wang2025explainability}. Consequently, models may still rely on spurious or clinically irrelevant features despite producing reasonable visual explanations.

In this work, we propose an \textit{expert-guided explainable few-shot learning} framework that directly incorporates radiologist-provided ROIs into the training process. Specifically, we introduce a novel explanation loss based on Dice similarity between Grad-CAM heatmaps and expert annotations, encouraging the model to focus on diagnostically meaningful regions during both training and inference. By integrating this spatial supervision with a prototypical network architecture, our approach enhances both classification accuracy and interpretability in low-data regimes.

We evaluate our framework on two clinically relevant datasets covering different modalities: (a) the BraTS dataset for brain tumor MRI classification and (b) the VinDr-CXR dataset for thoracic disease detection in chest X-rays. Experimental results demonstrate that our method consistently outperforms conventional few-shot baselines, while simultaneously producing more clinically meaningful attention maps that align with expert-defined diagnostic regions.

\section{Related Work}
FSL has emerged as a promising solution for training models with limited labeled data, a frequent challenge in medical imaging. Prototypical networks~\cite{snell2017prototypical,guo2024few} have become a widely adopted FSL approach due to their simplicity and effectiveness, learning class prototypes in an embedding space to perform classification based on distance metrics. In the medical domain, several extensions have been proposed to adapt few-shot models for tasks such as pathology image classification~\cite{quan2024dual} and multimodal fusion~\cite{ouahab2025protomed,guo2024few,chamarthi2024few}. However, these models primarily focus on improving classification accuracy while providing limited insight into model decision making processes, a critical requirement for clinical adoption.

XAI techniques have gained increasing attention for improving the transparency of deep learning models in healthcare~\cite{loh2022application,sadeghi2024review}. Among various methods, Grad-CAM~\cite{selvaraju2017grad,zhang2021grad} has become a widely used tool for generating post-hoc visual explanations by identifying salient regions that influence model predictions. Although XAI methods offer valuable interpretability, they are commonly employed after training and do not directly impact how models learn during optimization. Consequently, models may still exploit non-causal or clinically irrelevant features, even if their visual explanations appear reasonable.

Recent efforts have explored the integration of explanation supervision into model training to guide attention toward diagnostically relevant regions. For instance, Šefčík et al.~\cite{vsefvcik2023improving} utilized Layer-wise Relevance Propagation (LRP) to align model attention with glioma tumor regions in MRI, while Caragliano et al.~\cite{caragliano2025doctorintheloopexplainablemultiviewdeep} proposed the Doctor-in-the-Loop framework incorporating expert segmented regions for NSCLC CT analysis. Sun et al.~\cite{sun2021explanation} introduced explanation-guided training using LRP to improve cross-domain few-shot learning in natural images. However, these approaches either operate in full-data regimes, target non-medical domains, or lack direct integration with few-shot frameworks.

In contrast to prior work, our method introduces expert-guided attention alignment directly into few-shot learning for medical image diagnosis. By combining Grad-CAM-based spatial supervision with a prototypical network architecture, we encourage the model to focus on clinically meaningful regions during training, leading to improved generalization and interpretability under data-scarce conditions.

\section{Methodology}

We propose an expert-guided explainable few-shot learning framework that integrates expert-provided spatial annotations into the training process of a prototypical network. Our method combines few-shot classification loss with an explanation alignment loss based on Grad-CAM, encouraging the model to attend to diagnostically relevant regions.

\subsection{Few-Shot Learning with Prototypical Networks}

We adopt the prototypical network~\cite{snell2017prototypical} as the backbone for few-shot classification. Given an $N$-way $K$-shot classification problem, each episode consists of a support set $S = \{(x_s, y_s)\}_{s=1}^{N \cdot K}$ and a query set $Q = \{(x_q, y_q)\}_{q=1}^{N \cdot Q}$. A feature extractor $f_{\theta}(x)$, implemented as a DenseNet-121 model with its final classification layer removed, maps each image into a $d$-dimensional embedding space.

For each class $k$, a prototype $c_k$ is computed as the mean of the support embeddings:
\begin{equation}
c_k = \frac{1}{|S_k|} \sum_{(x_i, y_i) \in S_k} f_{\theta}(x_i).
\end{equation}
Classification for a query sample $x_q$ is performed based on the Euclidean distance between its embedding and the prototypes:
\begin{equation}
p(y=k \mid x_q) = \frac{\exp(-\|f_{\theta}(x_q) - c_k\|^2)}{\sum_{k'} \exp(-\|f_{\theta}(x_q) - c_{k'}\|^2)}.
\end{equation}
The prototypical classification loss is defined as:
\begin{equation}
\mathcal{L}_{\text{proto}} = -\log p(y = y_q \mid x_q)
\end{equation}

\subsection{Expert-Guided Attention Alignment via Grad-CAM}

To guide the model towards clinically meaningful features, we incorporate radiologist provided annotations during training. Specifically, Grad-CAM~\cite{selvaraju2017grad} is applied to generate attention heatmaps for query images, highlighting the spatial regions influencing model predictions.

Given an expert-provided binary mask $M$ corresponding to the ground truth abnormality for each query image, we define an explanation alignment loss using the Dice similarity coefficient between the Grad-CAM heatmap $G$ and the expert mask:
\begin{equation}
\mathcal{L}_{\text{exp}} = 1 - \frac{2 \cdot |G \cap M|}{|G| + |M|}.
\end{equation}
This loss penalizes misalignment between model attention and expert annotated diagnostic regions, encouraging the model to learn features that are both discriminative and clinically interpretable.

\subsection{Joint Optimization Objective}
The total training objective jointly optimizes both classification accuracy and attention alignment:
\begin{equation}
\mathcal{L}_{\text{total}} = \mathcal{L}_{\text{proto}} + \alpha \cdot \mathcal{L}_{\text{exp}},
\label{custom_loss}
\end{equation}
where $\alpha$ is a hyperparameter controlling the influence of explanation supervision. By optimizing $\mathcal{L}_{\text{total}}$, the model is encouraged to both correctly classify query samples and align its internal attention with expert-defined diagnostic regions, enhancing interpretability without sacrificing performance under limited data conditions.

\section{Experiments and Results}
In this section, we describe our experimental setup and present key findings. We analyze how varying the $\alpha$ coefficient influences classification performance and attention alignment. Additionally, we compare the accuracy and Grad-CAM focus maps of guided versus non-guided models to demonstrate the effectiveness of our explanation guided training approach in directing attention toward diagnostically relevant regions.

\subsection{Datasets}
We evaluate our expert-guided explainable few-shot learning framework on publicly available two expert-annotated medical imaging datasets spanning distinct modalities and annotation types.
\begin{itemize}
    \item \textbf{BraTS (MRI):} The BraTS dataset~\cite{menze2014multimodal} provides multimodal brain tumor MRI scans, including T1-weighted, T2-weighted, T2-FLAIR, and post-contrast T1 (T1Gd) sequences. Expert annotations are provided as segmentation masks delineating tumor subregions: edema dominant, necrotic dominant, and enhancing dominant. For our experiments, we stack T1Gd, T2, and T2-FLAIR images to form three-channel inputs, enabling the model to leverage complementary tissue contrasts.
     \item \textbf{VinDr-CXR (Chest X-ray):} The VinDr-CXR dataset~\cite{nguyen2022vindr} consists of frontal chest X-rays annotated by expert radiologists with bounding boxes marking thoracic abnormalities. We focus on three clinically significant findings: nodule/mass, pulmonary fibrosis, and lung opacity. Bounding boxes are converted into binary masks to serve as spatial supervision for attention alignment.

\end{itemize}

\subsection{Training Protocol}
We employ episodic training with a 3-way 3-shot few-shot classification setup. Each training episode consists of 9 support samples (3 per class) and 9 query samples, mimicking realistic low-data clinical scenarios. The feature extractor is initialized with DenseNet-121 weights. All models are trained for 7 epochs with 60 episodes per epoch. We compare two models:
\begin{itemize}
    \item \textbf{Non-guided baseline:} Prototypical network trained using $\mathcal{L}_{\text{proto}}$ only.
    \item \textbf{Guided model:} Our proposed method trained using $\mathcal{L}_{\text{total}}$ combining classification and explanation losses.
\end{itemize}
The explanation weight $\alpha$ is set to $0.10$ based on validation experiments.

\subsection{Evaluation Metrics}

We evaluate classification performance using overall accuracy and class-wise F1-scores, which better reflect performance across imbalanced diagnostic categories. To assess interpretability, we qualitatively compare Grad-CAM visualizations between guided and non-guided models to examine attention alignment with expert annotations.

\subsection{Performance Evaluation} 
We evaluated our explainability guided few-shot learning framework on two medical imaging datasets BraTS (MRI) and VinDr-CXR (X-ray), using both performance metrics and explainability analysis.

In quantitative evaluation, we compared guided and non-guided models using overall accuracy and class wise F1-scores. While accuracy provides a general performance overview, F1-score better reflects performance on each classes. These metrics help assess the model’s generalization across diagnostic categories as illustrated in Tables \ref{tab:model_accuracy} and \ref{tab:f1_score_comparison}.

\begin{figure}[tbp]
    \centering
    \includegraphics[width=0.90\columnwidth]{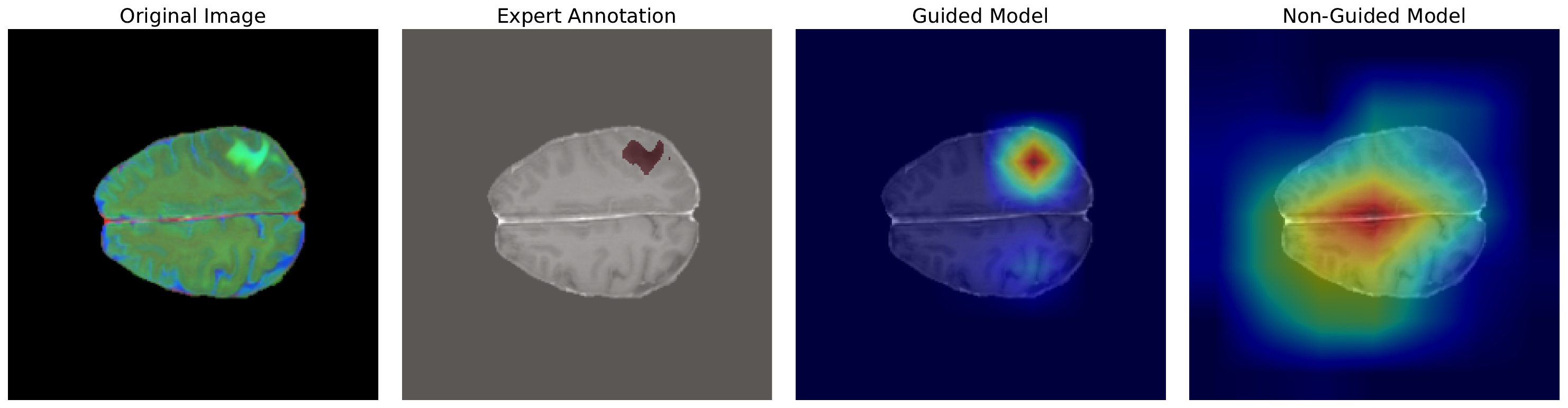}
    \includegraphics[width=0.90\columnwidth]{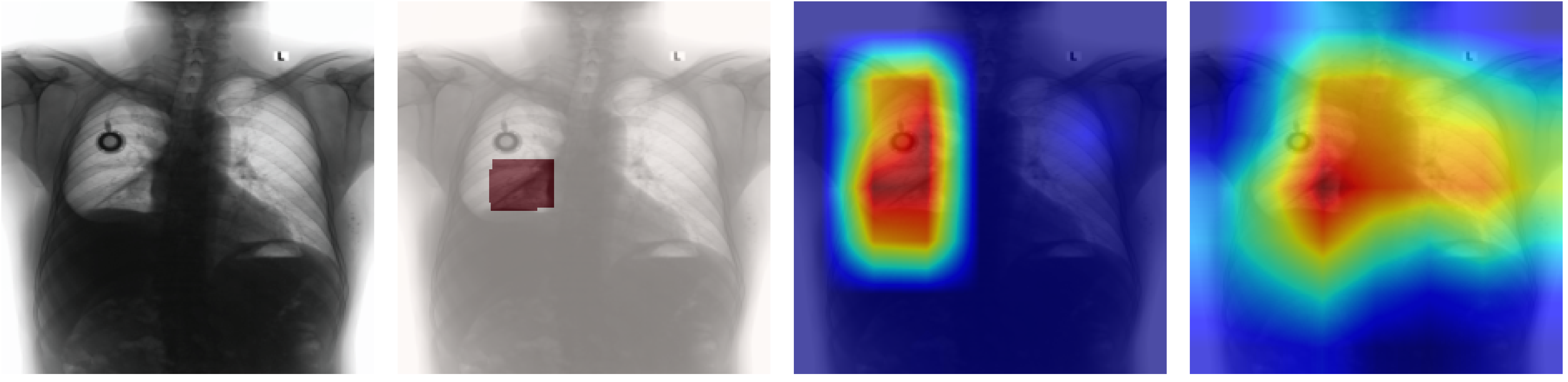}
    \caption{Grad-CAM heatmap comparison on BraTS (top) and VinDr-CXR (bottom). From left to right: Input Image, Expert Annotation, Guided Model Heatmap, Non-Guided Model Heatmap.}
    \label{fig:gradcam_comparison}
\end{figure}

On the other hand for explainability evaluation, we assess interpretability, where we used Grad-CAM to visualize model attention. By comparing heatmaps from guided and non-guided models, we evaluated whether attention aligns with expert annotated regions. The guided model consistently focused on clinically relevant areas, whereas the non-guided model often highlighted unrelated regions as shown in Figure \ref{fig:gradcam_comparison}.

\begin{table*}[tbp]
\centering
\caption{Comparison of classification {\em Accuracy} (in \%) between the guided and non-guided models across the BraTS and VinDr-CXR datasets, with varying training setups such as different training epochs and episodes per epoch, all using a fixed explanation weight $\alpha = 0.10$. The guided model consistently outperforms the non-guided baseline, demonstrating the benefit of incorporating expert-guided attention supervision.}
\label{tab:model_accuracy}
\begin{tabular}{|l|c|c|c|c|}
\hline
\textbf{Dataset} & \textbf{Epoch} & \textbf{Episodes} & \textbf{Guided Accuracy} & \textbf{Non-Guided Accuracy} \\
\hline
\multirow{3}{*}{\textbf{BraTS}} 
  & 7  & 60 & 83.61 & 77.09 \\
  & 10 & 30 & 69.41 & 68.12 \\
  & 15 & 20 & 72.14 & 68.49 \\
\hline
\multirow{3}{*}{\textbf{VinDr-CXR}} 
  & 7  & 60 & 73.29 & 50.65 \\
  & 10 & 30 & 67.26 & 54.33 \\
  & 15 & 20 & 54.47 & 51.19 \\
\hline
\end{tabular}%
\end{table*}

We assess the classification performance of our guided and non-guided models using two key metrics overall accuracy and class wise F1-score. Accuracy measures the proportion of correct predictions, while F1-score captures the balance between precision and recall, particularly important for clinical scenario. As shown in Table~\ref{tab:model_accuracy}, our guided model achieves higher accuracy on both datasets compared to our non-guided model from 77.06\% to 83.61\% on BraTS and from 54.33\% to over 73.29\% on VinDr-CXR. This indicates better generalization when attention alignment with expert annotations is introduced. Table \ref{tab:f1_score_comparison} presents the class wise F1-score comparison. The guided model consistently outperforms the non-guided model across all diagnostic categories, reflecting improved robustness and reliability. For example, in the VinDr-CXR dataset, F1-scores for Pulmonary Fibrosis and Nodule/Mass classes improved significantly under the guided setup.
These results demonstrate that integrating expert supervised attention not only enhances prediction accuracy but also yields more balanced and clinically trustworthy model performance.

\begin{figure}[tbp]
    \centering
    \includegraphics[width=0.90\columnwidth]{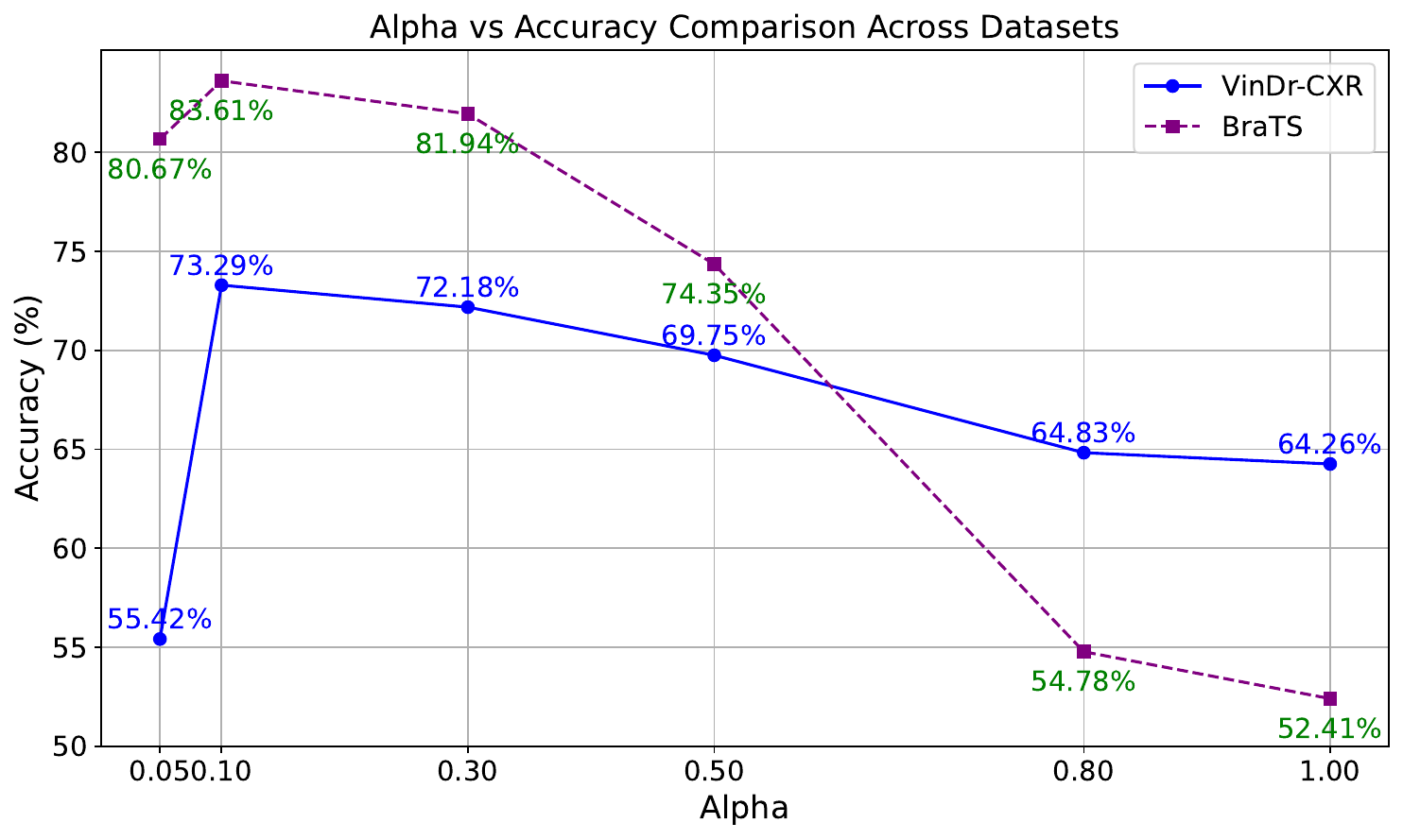}
    \caption{Accuracy comparison across different $\alpha$ values for both datasets}
    \label{fig:alpvavsacc}
\end{figure}

\subsection{Effect of Explanation Weight ($\alpha$)}
To analyze how explanation supervision influences model performance, we varied the explanation loss weight $\alpha$ in the loss function descibed in equation \eqref{custom_loss}. We evaluated multiple $\alpha$ values ranging from 0.05 to 1.0 on both the BraTS and VinDr-CXR datasets. As shown in Figure~\ref{fig:alpvavsacc}, the highest accuracy for both datasets was achieved at $\alpha = 0.10$, with BraTS reaching 83.61\% and VinDr-CXR reaching 73.29\%. Extremely low ($\alpha = 0.05$) or high ($\alpha = 1.0$) values led to performance drops, suggesting that both underemphasizing and overemphasizing explanation alignment can hinder learning. These results confirm that moderate explanation guidance helps the model learn more effectively while improving interpretability.

\subsection{Interpretability Analysis}
We qualitatively compare the guided and non-guided models using Grad-CAM visualizations. Figure~\ref{fig:gradcam_comparison} shows examples from the BraTS and VinDr-CXR datasets. The \textbf{non-guided model} often highlights irrelevant image regions, showing poor alignment with expert annotations. This undermines clinical trust, even when the classification accuracy is high.
In contrast, the \textbf{guided model}, trained with an attention alignment loss (Equation \eqref{custom_loss}), consistently focuses on diagnostically meaningful areas. This alignment with expert defined regions enhances the interpretability and trustworthiness of predictions, while also contributing to improved classification performance.

\subsection{Results Discussion}

Our experiments on the BraTS (MRI) and VinDr-CXR (X-ray) datasets show that incorporating expert guided explainability into few-shot learning significantly improves both performance and interpretability. The guided model achieved higher accuracy than the non-guided 83\% vs. 77\% on BraTS and 73\% vs. 54\% on VinDr-CXR dataset demonstrating better generalization in low data settings. Class wise F1 scores also improved, particularly for different classes on both dataset. We observed that the optimal explanation weight ($\alpha = 0.10$) provided the best balance between accuracy and alignment with expert annotations, too little or too much emphasis led to performance drops. Grad-CAM visualizations in figure \ref{fig:gradcam_comparison} confirmed that guided models focused on diagnostically relevant regions, unlike non-guided models which often highlighted irrelevant areas. These findings confirm that integrating expert supervision enhances both model performance and clinical interpretability.



\begin{table*}[tbp]
\centering
\caption{Class-wise F1 score (in \%) comparison between guided and non-guided models on the BraTS and VinDr-CXR datasets. The guided model consistently outperforms the non-guided counterpart across all classes, demonstrating better generalization.}
\label{tab:f1_score_comparison}
\resizebox{\textwidth}{!}{ 
\begin{tabular}{|c|c|c|c|}
\hline
\textbf{Dataset} & \textbf{Class} & \textbf{Guided Model F1 Score} & \textbf{Non-Guided Model F1 Score} \\
\hline
\multirow{3}{*}{\textbf{BraTS}} & Edema Dominant & 87.61 & 84.29\ \\
                                & Necrotic Dominant & 70.31 & 60.47 \\
                                & Enhancing Dominant & 66.95 & 64.23 \\
\hline
\multirow{3}{*}{\textbf{VinDr-CXR}} & Nodule/Mass & 72.55 & 57.03 \\
                                     & Pulmonary fibrosis & 70.20 & 35.60\\
                                     & Lung Opacity & 76.26 & 71.92 \\
\hline
\end{tabular}
}
\end{table*}

\section{Conclusion}
We proposed an expert-guided explainable few-shot learning framework that integrates radiologist-provided spatial annotations into model training to simultaneously improve classification accuracy and interpretability in medical image diagnosis. By introducing a novel explanation loss that aligns Grad-CAM attention maps with expert-defined regions, the model learns to focus on clinically meaningful features even under limited data conditions. Evaluations on both brain tumor MRI (BraTS) and chest X-ray (VinDr-CXR) datasets demonstrate consistent improvements in performance and visual interpretability compared to conventional few-shot baselines. Our findings suggest that incorporating expert-guided attention supervision offers a promising pathway toward more trustworthy and clinically deployable AI systems, particularly in data-constrained healthcare settings.

\section{Acknowledgment}
This work was supported in part by the National Science Foundation under Award \#2346643 and by the National Research Platform (NRP) Nautilus HPC cluster \cite{10.1145/3708035.3736060}.

\bibliographystyle{splncs04}
\bibliography{ref}

\end{document}